\documentclass[aps,prl,twocolumn,groupedaddress,showpacs,floatfix,superscriptaddress]{revtex4-1}
\usepackage{epsfig}
\usepackage{amsmath,amssymb}
\usepackage{graphicx}
\usepackage[dvipsnames,usenames]{color}
\usepackage[normalem]{ulem}
\usepackage{soul} 
\emergencystretch=\maxdimen
\begin{document}

\title{Charge Density Waves on a 
Half-Filled Decorated Honeycomb Lattice}
\author{Chunhan Feng} 
\affiliation{Department of Physics, University of California, 
Davis, CA 95616, USA}
\author{H. Guo}
\affiliation{Department of Physics, Key Laboratory of Micro-Nano
Measurement-Manipulation and Physics (Ministry of Education),
Beihang University, Beijing, 100191, China}
\author{R.T. Scalettar}
\affiliation{Department of Physics, University of California, 
Davis, CA 95616, USA}

\begin{abstract}
Tight binding models like the Hubbard Hamiltonian are most often
explored in the context of uniform intersite hopping $t$.  The
electron-electron interactions, if sufficiently large compared to this
translationally invariant $t$, can give rise to ordered magnetic phases
and Mott insulator transitions, especially at commensurate filling.  The
more complex situation of non-uniform $t$ has been studied within a
number of situations, perhaps most prominently in multi-band geometries
where there is a natural distinction of hopping between orbitals of
different degree of overlap.  In this paper we explore related questions
arising from the interplay of multiple kinetic energy scales and {\it
electron-phonon} interactions.  Specifically, we use Determinant Quantum
Monte Carlo (DQMC) to solve the 
half-filled Holstein Hamiltonian on a `decorated
honeycomb lattice', consisting of hexagons with internal hopping $t$
coupled together by $t^{\,\prime}$.  This modulation of the hopping
introduces a gap in the Dirac spectrum and affects the nature of the
topological phases.  We determine the range of $t/t^{\,\prime}$ values
which support a charge density wave (CDW) phase about the Dirac point of
uniform hopping $t=t^{\,\prime}$, as well as the critical transition
temperature $T_c$.  The QMC
simulations are compared with the results of Mean Field Theory (MFT).
\end{abstract}

\date{\today}

\maketitle

\section{1.  Introduction}

Itinerant electrons on a honeycomb lattice host a Dirac spectrum in the
absence of interactions which has attracted considerable
attention \cite{haldane88,kane05,novoselov05,li07,wunsch08,wehling14}.
The linearly vanishing density of states (DOS) at $E=0$ forms an
interesting counterpoint to that of the square lattice (of interest to
cuprate superconductivity) whose DOS
diverges (logarithmically) at $E=0$.  An immediate consequence is that,
whereas in the square lattice long range antiferromagnetic (AF)
correlations onset in the ground state for any finite repulsive
interaction $U$, a nonzero critical $U_c$ is required for AF  order on
the honeycomb lattice \cite{paiva05,sorella12}.

Recently, the effects of electron-phonon interactions on Dirac fermions
have been explored \cite{zhang19,chen19}.  Similar to the case of
electron-electron interactions, the semi-metallic band structure requires 
a critical electron-phonon interaction
strength for CDW formation at half-filling.  A crucial
difference is that, unlike Ne\'el order which occurs only at $T=0$ in
the two dimensional Hubbard model \cite{hirsch89}, owing to the
continuous nature of the spin symmetry being broken, the CDW transition
occurs at finite temperature. 

In this paper we extend these investigations of the Holstein model on a
honeycomb lattice by examining the effect of a regular pattern of
non-uniform hopping.  The particular `Kekul\'e hopping texture' we
investigate has been proposed \cite{wu15} to give rise to nontrivial
topological properties associated with an opening of a gap at the Dirac
point, and linked to the `pseudo-angular momentum' of electrons residing
on sets of strongly hybridized hexagons.  Similar `decorated lattices'
have been studied previously in the context of the depleted square
lattice Heisenberg \cite{troyer96} and Hubbard \cite{khatami14}
Hamiltonians as possible theoretical descriptions of spin liquid phases
in CaV$_4$O$_9$ \cite{taniguchi95,katoh95,ueda96,gelfand96,pickett97}.
It was shown that while long range antiferromagnetic correlations exist
in the ground state when the hoppings $t$ and $t^{\,\prime}$ are roughly
balanced, spin liquid phases consisting of independent spin dimers or
spin plaquettes are present when the hoppings are sufficiently unequal.
Within mean field theory, a rich variety of spin-ordered phases,
characterized by different patterns of spin inside and between the
plaquettes, can arise as a function of doping and $U$ in such decorated
Hubbard models \cite{khatami14}.

Strongly correlated physics in the presence of several kinetic energy
scales gives rise to a further variety of phenomena in other important
realizations, including orbitally-selective Mott transitions
\cite{liebsch05,arita05,inaba06,koga04,ferrero05,demedici05,biermann05}.
In the case of the Periodic Anderson Model (PAM) which includes both
conduction $c$ and local $d$ orbitals, a dominant interorbital hopping
$t_{cd} \gtrsim t_{cc}$ can lead to singlet formation and a spin-liquid
ground state \cite{lee86,nishino93}, as seen in QMC studies in $d=1,2,3$
and $d=\infty$ \cite{fye90,jarrell93,vekic95,huscroft99}.  As in the
less widely studied case of decoration, the existence of several hopping
energy scales whose difference is large disrupts magnetic order.  Most
of these investigations have focused on electron-electron interactions.

We will discuss some interesting analogies between the spin-singlet
formation in such situations, and charge singlets in the electron-phonon
case.
However, it is important to emphasize that the Holstein model breaks the
spin symmetry present in the Hubbard model, with charge order (the
analog of ordering in the $S^z$ channel) dominating over superconducting
order (which maps onto $S^x, S^y$).  As a result,
the CDW transition in the half-filled 2D Holstein model occurs at finite
temperature \cite{weber18}, whereas long range magnetic order in the 2D Hubbard model
occurs only at $T=0$.  This breaking of symmetry introduces a
fundamental difference between the physics of the repulsive Hubbard and
Holstein models with multiple hopping energy scales, which is especially
marked as the phonon frequency $\omega_0$ decreases.

This paper is organized as follows: Section 2 introduces the precise
model we will investigate, along with our computational methodology.
Section 3 discusses the results of mean field theory (MFT) calculations,
which we show capture some of the tendency to reduced CDW order with
nonuniform hoppings.  Section 4 contains the detailed
DQMC results and analysis, and is followed by some further discussion
and interpretation in Sec.~5.

\begin{figure}[t]
\includegraphics[height=6.0cm,width=8.0cm]{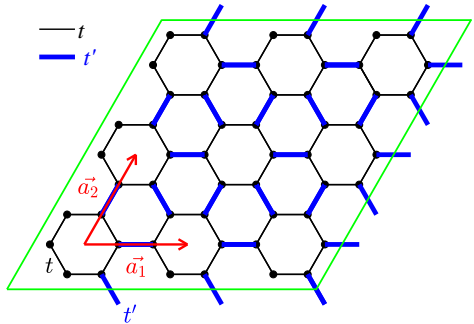}
\caption{
The structure of the `decorated honeycomb lattice'.
Two different hopping strengths are present.  Hybridization $t$ (thin
black lines) links the sites of a collection of independent hexagons.
These hexagons are then connected by $t^{\,\prime}$ (thick blue lines).
In the $t^{\,\prime} >> t$ limit, an alternate description in terms of
elemental  dimers linked by $t$ is a more appropriate starting point.
}
\label{fig:lattice}
\end{figure}

\section{2. Model and Methods}

We investigate the Holstein Hamiltonian,
\begin{align}
\hat{\cal H}=
&- \sum_{\bf \langle ij \rangle} t_{\bf \rm ij}
\big(
\hat c^{\dagger}_{{\bf i}\sigma}
\hat c^{\phantom{\dagger}}_{{\bf j}\sigma}
+ \hat c^{\dagger}_{{\bf j}\sigma}
\hat c^{\phantom{\dagger}}_{{\bf i}\sigma}
\big)  
-\mu \sum_{{\bf i}\,\sigma} n_{{\bf i}\,\sigma}
\nonumber \\
&+ \frac{1}{2} \sum_{\bf i} \hat p_{\bf i}^2 +
\frac{1}{2} M \omega_0^2 \sum_{\bf i} \hat x_{\bf i}^2 +
\lambda \sum_{\bf i} \hat x_{\bf i}
(\hat n_{{\bf i}\uparrow} + \hat n_{{\bf i}\downarrow})
\,\,.
\label{eq:ham}
\end{align}
Here the kinetic energy sum is over sites on a hexagonal lattice, with
$t_{\bf \rm ij}=t$ for pairs of sites internal to a set of hexagons, and
$t_{\bf \rm ij}=t^{\,\prime}$ for pairs of sites bridging distinct
hexagons.  See Fig.~\ref{fig:lattice}.  We will report lattice sizes $N$
in terms of the number of hexagons, i.e.~the unit cell count.
Figure~\ref{fig:lattice} corresponds to $N=(3 \times 3) \times 6=54$
sites.  The remaining terms in $\hat {\cal H}$ consist of a collection
of local quantum oscillators of frequency $\omega_0$ ($\omega_0=1$ is
used in all the simulations in this paper) and an electron-phonon
coupling $\lambda$ of the fermionic charge density $\hat n_{{\bf
i}\uparrow} + \hat n_{{\bf i}\downarrow}$ to the displacement $\hat
x_i$.  We will measure the strength of the coupling via the
dimensionless combination $\lambda_D \equiv 
\lambda^2/(M\omega_0^2 \, W)$.  
In the anti-adiabatic limit
$\omega_0 \rightarrow \infty$,
the coupling $\lambda_D$ can be thought of as the ratio of an
effective attraction between electrons mediated by the phonons, $U_{\rm
eff} = -\lambda^2/(M \omega_0^2)$, to the kinetic energy scale $W$.  The
choice $t=1+\Delta$ and $t^{\,\prime}=1-2\Delta$ keeps the bandwidth
$W=6$ fixed as $\Delta$ is varied, allowing us to
study the effects of modulated hopping while keeping 
$\lambda_D$ constant.
We set the phonon mass $M=1$ and
tune the chemical potential  $\mu= -\lambda^2/\omega_0^2$ 
to the particle-hole symmetry point
so that the filling is always 
$\langle \, n_{{\bf i}\,\sigma} \, \rangle=\frac{1}{2}$.

We solve for the properties of Eq.~\ref{eq:ham}
using two methods.  The first is a mean field approach in
which we make an {\it ansatz} for the phonon coordinates.
(See Sec.~3.)
The resulting Hamiltonian is quadratic in the remaining
fermion degrees of freedom and can be solved analytically.
The free energy is minimized within the parameter space allowed in the
{\it ansatz}.  The second approach is DQMC\cite{blankenbecler81,white89}.  
Unlike MFT, it solves the many-body problem exactly, 
on finite lattices.
DQMC has statistical errors associated with the sampling, which 
are of the order of 0.1\% for local quantities like the
double occupancy and energy, but can be several percent
for global quantities like structure factors in the vicinity
of phase transitions.  DQMC also has 
`Trotter errors'\cite{trotter59,suzuki76,fye86,fye87}
arising from the discretization of imaginary time.
Because these Trotter errors are of the same order as, or smaller than, 
the statistical ones
for the quantities we use in determining the phase boundary, we do not
perform any extrapolation in the imaginary time discretization.
\color{black}
For all the work in this paper we use $\Delta\tau=0.1$.
\color{black}

Regardless of the value of $\Delta$, the decorated honeycomb lattice is
bipartite, and hence the local fermionic pairs which form due to the
effective attractive interaction $U_{\rm eff}$ mediated by the phonons
tend to form a charge density wave phase at half-filling.  Previous
investigations have determined the phase diagram in the $\lambda_D$-$T$
plane for $t=t^{\,\prime}$ \cite{zhang19,chen19}.  For $\omega_0/t=1$,
there is a quantum critical point at $(\lambda_D)_c=0.27$ above which
CDW order forms in the ground state.  $T_c$ rises rapidly at
$(\lambda_D)_c$, reaching a maximum value $T_c/t \sim 0.2$ at $\lambda_D
\sim 0.5$.  We are interested here in the effect of the
nonuniform hoppings $\Delta$ on $T_c$ and on
$(\lambda_D)_c$.  In the limits $\Delta=0.5$ and $\Delta=-1$ the system
separates into collections of independent hexagons and dimers, making
long range order impossible and $T_c=0$ trivially.

\begin{figure}[t]
\includegraphics[height=9.0cm,width=8.0cm]{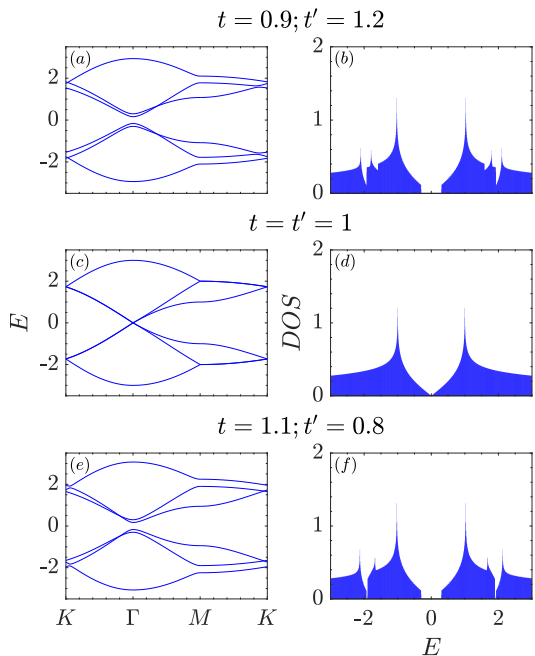}
\caption{
Left: Energy dispersion $E(\vec{k})$ in the noninteracting ($\lambda=0$)
limit.  The Dirac points of the two bands of the honeycomb lattice,
$t=t^{\,\prime}$ (middle panel), are split by the decoration $t \neq
t^{\,\prime}$.  In both cases, $t>t^{\,\prime}$ and $t<t^{\,\prime}$, a
gap is opened at half-filling.  See text for a discussion of differences
at other fillings.  Right: Density of states for the same three cases as
left panel.  A gap at half-filling is evident when $t \neq
t^{\,\prime}$.
}
\label{fig:energy_spectrum}
\end{figure}

The two site unit cell of the honeycomb lattice is expanded by the
decoration, so that now there are six bands.  Figure
\ref{fig:energy_spectrum} shows $E(\vec{k})$ for the undecorated
honeycomb lattice $t=t^{\,\prime}$ (central panel); the dimer limit
$t<t^{\,\prime}$ (top panel); and the hexagon limit $t^{\,\prime}<t$
(bottom panel).  In either case, the touching of the
two bands at the Dirac cones which occurs at half-filling and
$t=t^{\,\prime}$ is replaced by a gap.

The associated densities of states (DOS) for the three cases are shown
in Fig.~\ref{fig:energy_spectrum} Right.  Consistent with the dispersion
relations of Fig.~\ref{fig:energy_spectrum} Left, when $\Delta \neq 0$,
the linearly vanishing DOS at $E=0$ of the isotropic honeycomb lattice
is replaced with a gap. 

The  decorated lattice geometry of Fig.~\ref{fig:lattice} has been
proposed as a generalization of the isotropic honeycomb lattice with a
topological gap opened by the difference between the inter- and
intra-plaquette hoppings \cite{wu15}.  The six resulting bands can be
viewed as arising from the six single electron states (``orbitals")
which exist on each independent ($t^{\,\prime}=0$) hexagon and whose
degenerate levels are broadened when $t^{\,\prime}\neq 0$.  The
topological nature is not like that induced by spin-orbit coupling.
Instead, it is similar to the 1D Su-Schrieffer-Heeger model, which also
contains weak and strong bonds.  Domain walls which arise from $t \neq
t^{\,\prime}$ are associated with a gapless boundary state. 

Other versions of decoration exist.  For example, R\"uegg {\it etal}
\cite{ruegg10} have explored topological insulators of a tight-binding
Hamiltonian with spin-orbit and Rashba interactions on a ``star" lattice
which interpolates between honeycomb and Kagom\'e geometries.
Similarly, when honeycomb rhodates like Li$_2$RhO$_3$, are pressurized
various bond dimerization patterns emerge on the Rh hexagons, and are
associated with different magnetic patterns \cite{hermann19}.  A final
example is strained graphene, in which the hoppings $t_1, t_2, t_3$
along the three primitive lattice vectors are allowed to be
unequal \cite{sharma12,sorella18,gamayun17,andrade19}.

As noted in the introduction, in quantum spin-1/2 and itinerant electron
Hamiltonians with repulsive interactions, unequal hoppings tend to
degrade long range magnetic order.  It is worth discussing the relation
between those (spin) singlet phases and the disordered phases in the
attractive Hubbard model, since that has a close connection to the
Holstein model studied here; both exhibit CDW and superconducting phases
and a quantitative link is provided by $U_{\rm eff} =
-\lambda^2/\omega_0^2$.

In particular,
consider the well-known particle-hole transformation (PHT)
$c^{\dagger}_{{\bf i}\downarrow} \rightarrow (-1)^{\bf i}
c^{\phantom{\dagger}}_{{\bf i}\downarrow}$
on the down spin fermions.  
On a bipartite lattice, and at $\mu=0$, this PHT leaves the kinetic 
energy unchanged, but reverses the sign of the interaction term.
The different components of the spin operator
map into charge and pairing correlations,
\begin{align}
S_{\bf i}^z &\equiv 
n^{\phantom{z}}_{{\bf i} \uparrow} - n^{\phantom{z}}_{{\bf i} \downarrow}
\hskip0.05in \rightarrow \hskip0.05in
n^{\phantom{z}}_{\bf i} \equiv 
n^{\phantom{z}}_{{\bf i} \uparrow} + n^{\phantom{z}}_{{\bf i} \downarrow}
\nonumber \\
S^{+}_{\bf i} &\equiv 
c^{\dagger}_{{\bf i}\uparrow}
c^{\phantom{\dagger}}_{{\bf i}\downarrow}
\hskip0.27in \rightarrow \hskip0.05in
\Delta^{\dagger}_{\bf i} \equiv
(-1)^{\bf i} \, c^{\dagger}_{{\bf i}\uparrow}
c^{\dagger}_{{\bf i}\downarrow}
\nonumber \\
S^{-}_{\bf i} &\equiv 
c^{\dagger}_{{\bf i}\downarrow}
c^{\phantom{\dagger}}_{{\bf i}\uparrow}
\hskip0.27in \rightarrow \hskip0.05in
\Delta^{\phantom{\dagger}}_{\bf i} \equiv
(-1)^{\bf i} \, c^{\phantom{\dagger}}_{{\bf i}\downarrow}
c^{\phantom{\dagger}}_{{\bf i}\uparrow}
\label{eq:pht}
\end{align}

This PHT yields insight into some of the expected physics in the
presence of attractive interactions.  In analogy with the formation of
spin singlets in the repulsive case, for the attractive Hubbard and
Holstein models we expect the development of `charge singlets' in which
the three components of charge/pairing operators on the right side of
Eq.~\ref{eq:pht} form local objects on either dimers or hexagons.  These
charge singlets might then compete with long range CDW order when $t$ and
$t^{\,\prime}$ differ too greatly.  

\begin{figure}[t]
\includegraphics[height=6.0cm,width=8.0cm]{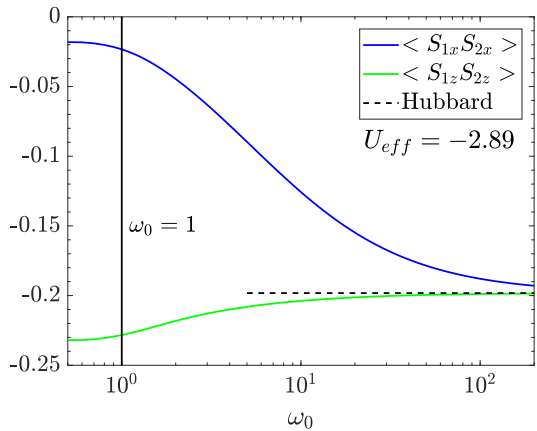}
\caption{
Exact diagonalization results for the ground state
charge, $\langle \, S^z_1 S^z_2 \, \rangle$, and superconducting,
$ \langle \, S^x_1 S^x_2 \, \rangle$, 
correlators (in magnetic language) on a two site Holstein dimer, as a function of
$\omega_0$ at fixed $U_{\rm eff}=-2.89$.  
The vertical line at $\omega_0=1$ shows the phonon
frequency used
in the phase diagram obtained in this paper.
The dashed horizontal line is the Hubbard model
result at $U=2.89$.
\textcolor{black}{It is notable that values 
$\omega_0/t \lesssim 1$ are quite far from the 
limit where the spin correlations (CDW-pairing correlations)
are symmetric.}
}
\label{fig:dimer}
\end{figure}

With this said, it is worth emphasizing
that the Holstein $\, \leftrightarrow \,$ Hubbard mapping is
exact only in the anti-adiabatic limit
$\omega_0 \rightarrow \infty$.
Figure~\ref{fig:dimer} shows
the effect of finite phonon frequency $\omega_0$
on the different components of Eq.~\ref{eq:pht}.
Symmetry is restored as $\omega_0 \rightarrow \infty$, but for the
value $\omega_0=1$ used in this paper, 
$\langle \, S^z_1 S^z_2 \, \rangle \ll
 \langle \, S^x_1 S^x_2 \, \rangle$.
Thus, while the analogy to magnetic physics is useful,
it is far from clear how it will manifest itself quantitatively.
(The fact that these correlators are less in magnitude
than the singlet value $-1/4$ is due to charge fluctuations.
As $U$ also becomes large, they approach the 
Heisenberg limit $-\frac{1}{4}$ so that $\vec S_1 \cdot \vec S_2
=-\frac{3}{4}$.)

\section{3.  Mean Field Results}

We first examine the physics of the Hamiltonian of 
Eq.~\ref{eq:ham} within mean field theory.  In this approach,
we ignore the phonon kinetic energy and assume a staggered pattern
for the phonon displacements, $x_i = x_0 + (-1)^i \, x_1$.
Here $(-1)^i=\pm 1$ on the two sublattices of the (bipartite)
honeycomb geometry.  The quadratic fermion Hamiltonian
can be diagonalized, resulting in a 
total free energy per site
which combines both electron and phonon contributions,
$f(x_0,x_1,T)
=N \, w_0^2 \, (x_0^2 + x_1^2)/2 
- T \, \sum_{\alpha} {\rm ln} \big( \, 1 + e^{-\epsilon_\alpha(x_0,x_1)/T} \, \big)$,
where $\epsilon_\alpha$ are the fermion energy levels.  A nonzero bond
dimerization $x_1$ implies an associated charge modulation, since
$\lambda x_i$ acts as a local chemical potential on site $i$.

\begin{figure}[t]
\includegraphics[height=6.0cm,width=8.0cm]{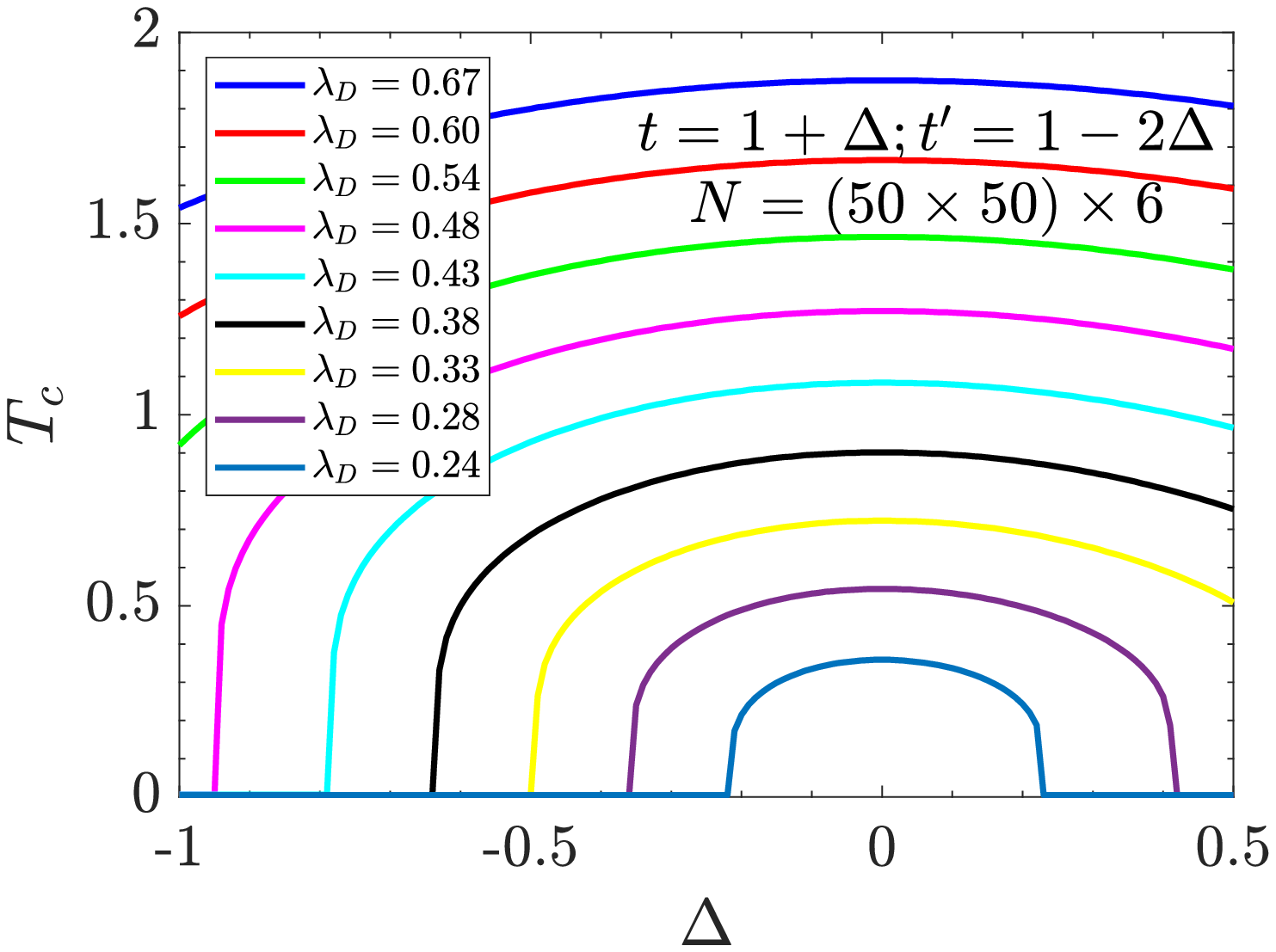}
\includegraphics[height=6.0cm,width=8.0cm]{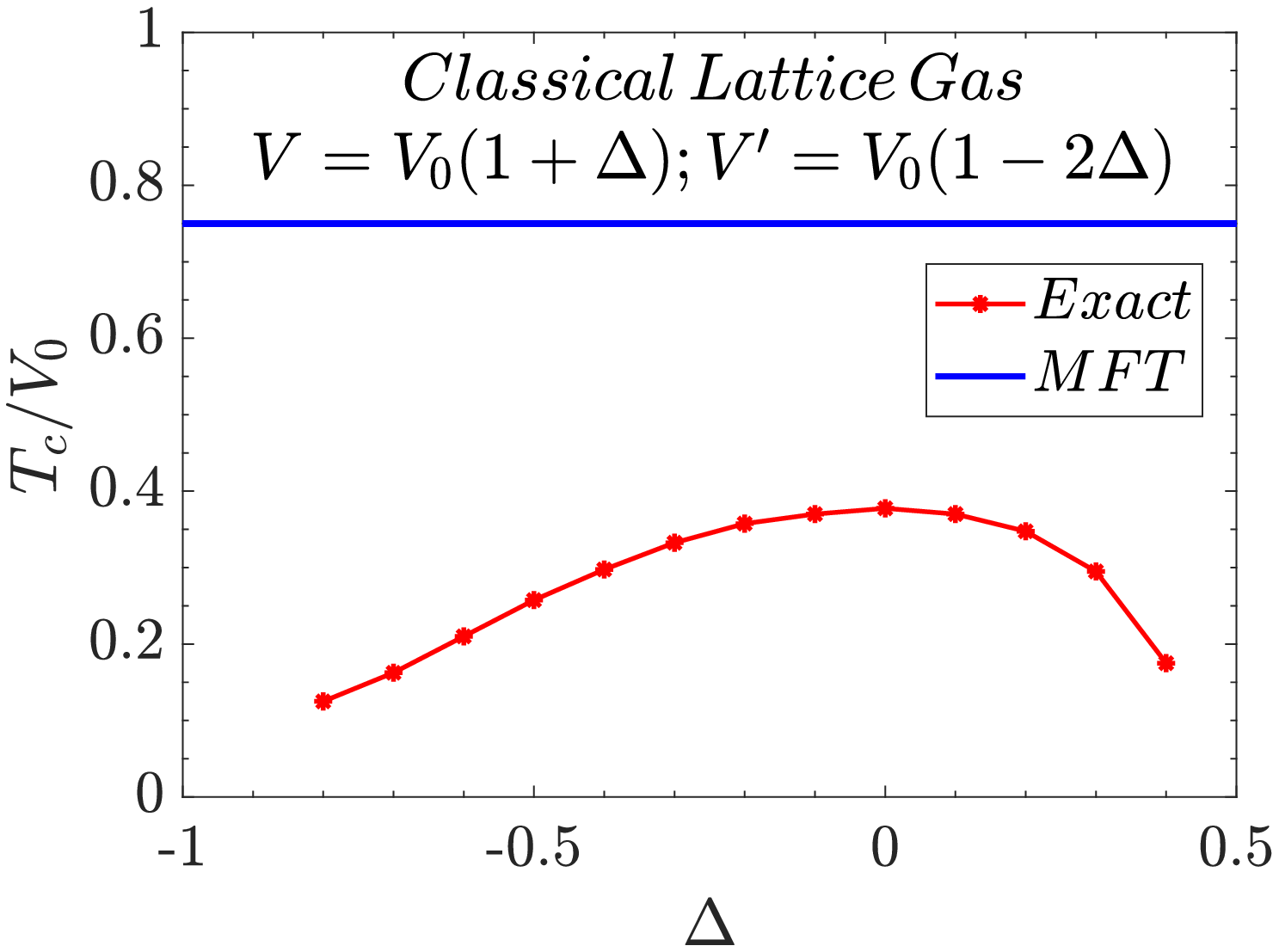}
\caption{
Top: 
Dependence of $T_c$ on $\lambda_D$
of the decorated Holstein model Eq.~\ref{eq:ham} 
within mean field theory.  
The CDW transition temperature is maximized for isotropic hopping
($\Delta=0$), and is suppressed on the dimer side $\Delta<0$ and the
hexagon side $(\Delta>0$).  Bottom: Comparison of $T_c$
given by MFT and classical Monte Carlo for a classical lattice gas.  The
difference between exact $T_c$ and MFT $T_c$ is more significant when
$\Delta$ approaches to the limiting cases $\Delta=-1$ and $\Delta=0.5$.
}
\label{fig:MFT_L4lambda2}
\end{figure}

The resulting phase diagram is shown in the top panel of
Fig.~\ref{fig:MFT_L4lambda2}.  $T_c$ is decreased by decoration, as
might be expected from the Stoner criterion and the opening of a true
gap (vanishing of the Fermi surface density of states in a finite
chemical potential range).  However, for $\lambda_D=\frac{2}{3}$, the effect is
relatively small:  Even in the extreme independent hexagon and dimer
limits $T_c(\Delta=0.5)/T_c(\Delta=0) = 0.965 $ and
$T_c(\Delta=-1)/T_c(\Delta=0) =0.822 $, respectively.   The MFT $T_c$ is
nonzero even though there can be no symmetry breaking on small finite
clusters. On the other hand, for smaller $\lambda_D$, MFT results indicate
that a critical $\Delta$ is needed in order to have a CDW phase, which
is consistent with the DQMC results in
Fig.~\ref{fig:Tc_Delta_phase_diagram}.
We have verified that, for $\lambda_D>0.24$,
the MFT results shown for $(50\times50)\times6$ lattices
change by less than the thickness of the lines
if the lattice size is decreased to $(4\times4)\times6$, an observation
which aids in  interpreting the
DQMC results of the next section, which are necessarily on smaller lattices.
At $\lambda_D<0.24$, where $T_c$ gets small and the CDW region is
minute, finite size effects, unsurprisingly, become more pronounced.

It is interesting to contrast this with the behavior of the simplest
model of CDW physics in this geometry, the classical lattice gas $E=
\sum_{\langle ij \rangle} V_{ij} n_i n_j$.  Here $n_i =0, 1$ and we
choose $V_{ij}=V_0 (1 + \Delta)$ or $V_{ij}=V_0 (1 - 2\Delta)$, with the
same geometry and bond convention as in Fig.~\ref{fig:lattice}.  The
total coupling $\sum_j V_{ij}$ at each site $i$ is independent of
$\Delta$, in analogy to fixing the bandwidth $W$.  The transition
temperature as a function of $\Delta$ is given in the bottom panel of
Fig.~\ref{fig:MFT_L4lambda2}.  Within MFT, $T_c$ is completely
independent of $\Delta$ because $T_c$ is only a function of the total,
and invariant, $\sum_j V_{ij}$.  The bottom panel of
Fig.~\ref{fig:MFT_L4lambda2} also gives the exact $T_c$ (obtained by
Binder crossings of classical Monte Carlo simulations).  The exact $T_c$
does depend on $\Delta$, and can be seen to vary by a factor of three
from its $\Delta=0$ value when $\Delta =-0.8$ or $\Delta=+0.4$, values
which approach the decoupled hexagon and dimer limits.  Unlike the MFT
calculation, the exact $T_c$ must vanish at $\Delta=-1$ and
$\Delta=+0.5$, and the lattice consists of independent clusters.

As we shall see in the following section, 
the MFT values for the critical temperature
of Fig.~\ref{fig:MFT_L4lambda2}(top) are an order of magnitude larger
than those of QMC.  This is perhaps not too surprising given
the low dimensionality being studied.  We note that a
similar comparison of phase diagrams for the 2D Hubbard
model revealed MFT in considerable disagreement with DQMC\cite{hirsch85}.

\section{4.  Quantum Monte Carlo Results}

We now turn to the results of DQMC simulations which include
fluctuations neglected in the preceding MFT treatment.
We begin by showing the charge structure factor,
\begin{align}
S_{\rm cdw} = \frac{1}{N} \sum_{i,j} \, (-1)^{i+j} \,
\langle \, n_i n_j \, \rangle.
\label{eq:sdw}
\end{align}
with $(-1)^{i+j} = \pm 1$ according to whether sites $i,j$ are on the
same or different sublattices.  In an ordered phase, $T<T_c$, we expect
$S_{\rm cdw}$ to grow linearly with the lattice size since $\langle \,
n_i n_j \, \rangle $ is non-zero even for widely separated $i,j$ pairs.  

\begin{figure}[t]
\includegraphics[height=6.0cm,width=8.0cm]{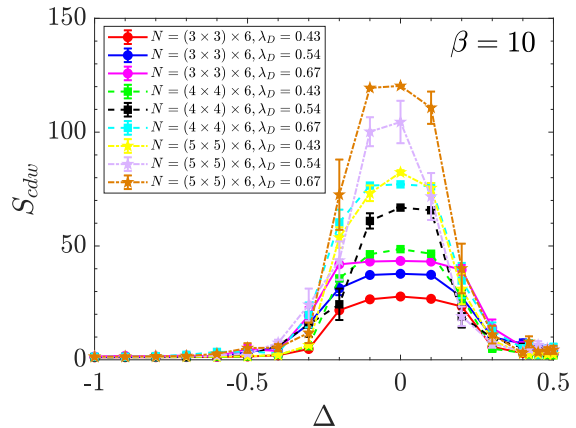}
\caption{
Charge structure factor $S_{\rm cdw}$ as a function of
hopping difference $\Delta$.  There is a window
near the isotropic point $\Delta=0$ in which $S_{\rm cdw}$
is large and scales with system size, indicating long range charge
order.
}
\label{fig:Scdw_delta}
\end{figure}

Figure \ref{fig:Scdw_delta} gives $S_{\rm cdw}$ for several values of
$\lambda_D$ and lattice sizes $N$ at low temperature, $\beta= 10$.  In a
window about the isotropic Holstein limit ($\Delta=0$), $S_{\rm cdw}$ is
large and increases with lattice size, suggesting the presence of long
range charge correlations for those values.  Meanwhile, for large
$\Delta$, $S_{\rm cdw}$ is small and independent of
size.  Two quantum critical points (QCP) $\Delta_c$ separate the CDW
from charge singlet regions at the two extremes of
hopping difference $\Delta=-1.0$ and.  $\Delta=0.5$.

\begin{figure}[t]
\includegraphics[height=6.0cm,width=8.0cm]{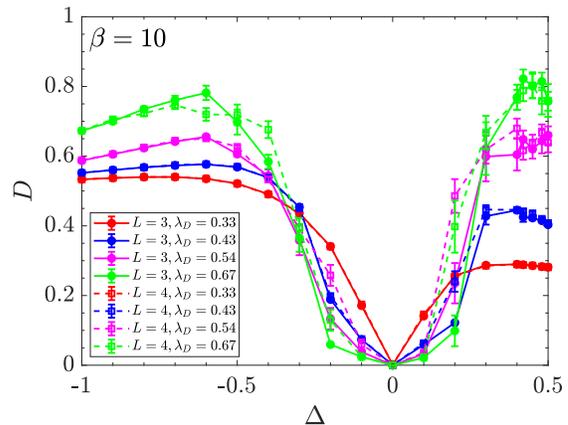}
\caption{
Difference ${\cal D}$ between density correlation
function on $t$ and $t^{\, \prime}$ bonds as a function of
$\Delta$.  ${\cal D}$ rises steeply in the vicinity
of the CDW to charge singlet QCP.
}
\label{fig:difference_ddcorr}
\end{figure}

The difference
\begin{align}
{\cal D} \equiv 
C_{\rm nn} - C_{\rm nn}^{\,\prime}  
\equiv
\langle \, n_i n_{i+\hat x} \rangle_{t}
          -   \langle \, n_i n_{i+\hat x} \rangle_{t^{\,\prime}}
\end{align}
between the near-neighbor density-density correlations $C_{\rm nn}$ on
the $t$ and $t^{\,\prime}$ bonds provides a measure of the effect of
hopping difference on the order.  For $\Delta$ small,
${\cal D}$ is small.  ${\cal D}$ rises rapidly in the vicinity of the
QCPs $\Delta_c$.  See Fig.~\ref{fig:difference_ddcorr}.  Indeed, $d{\cal
D}/d\Delta$ can be regarded as a ``inhomogeneity susceptibility" which
diverges at $T=0$ as $\Delta \rightarrow \Delta_c$.

\begin{figure}[b]
\includegraphics[height=6.0cm,width=8.0cm]{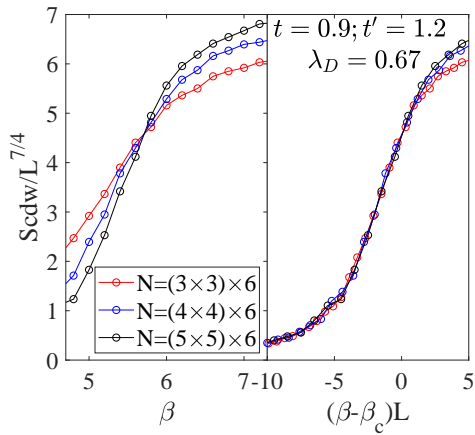}
\caption{
Scaled structure factor $S_{\rm cdw}/L^{\gamma/\nu}$ as a function of
$\beta$.  The scaling exponent $\gamma/\nu=7/4$ is taken to be the 2D
Ising value, and provides a good universal crossing.  The crossing
points identify $T_c = 1/\beta_c$.
}
\label{fig:bindercross}
\end{figure}
Figures \ref{fig:Scdw_delta}--\ref{fig:difference_ddcorr} focus on the
low temperature charge correlations and identify the positions of the
QCPs which bound the CDW regions near the isotropic lattice limit.
Within the CDW, there is a finite temperature phase transition as $T$ is
decreased.  Crossings of the scaled structure factor $
S_{\rm cdw}/L^{\gamma/\nu}$ are shown in Fig.~\ref{fig:bindercross} and identify $T_c$
in the region of small $\Delta$ where long range order persists.

As discussed in the introduction, it seems natural to connect the loss
of charge order in this electron-phonon model to analogous AF-singlet
transitions in Hamiltonians describing quantum magnetism which have
several exchange energy scales, e.g.~the periodic Anderson and bilayer
Hubbard Hamiltonians, and the bilayer or random bond Heisenberg
Hamiltonians.  We have used that language extensively
in the present paper, since it does constitute
a useful touchstone.  However, the fact that a finite temperature CDW
transition occurs in the Holstein model suggests care should be taken in
emphasizing this connection, since the continuous symmetry of the
ordering direction forbids such a finite $T$ transition in the 2D
magnetic models.  
Indeed, Figure \ref{fig:dimer} shows that the parameters explored in
Figs.~\ref{fig:Scdw_delta}-\ref{fig:bindercross}
are in fact very far from the regime where the analogy is precise.

\begin{figure}[t]
\includegraphics[height=6.0cm,width=8.0cm]{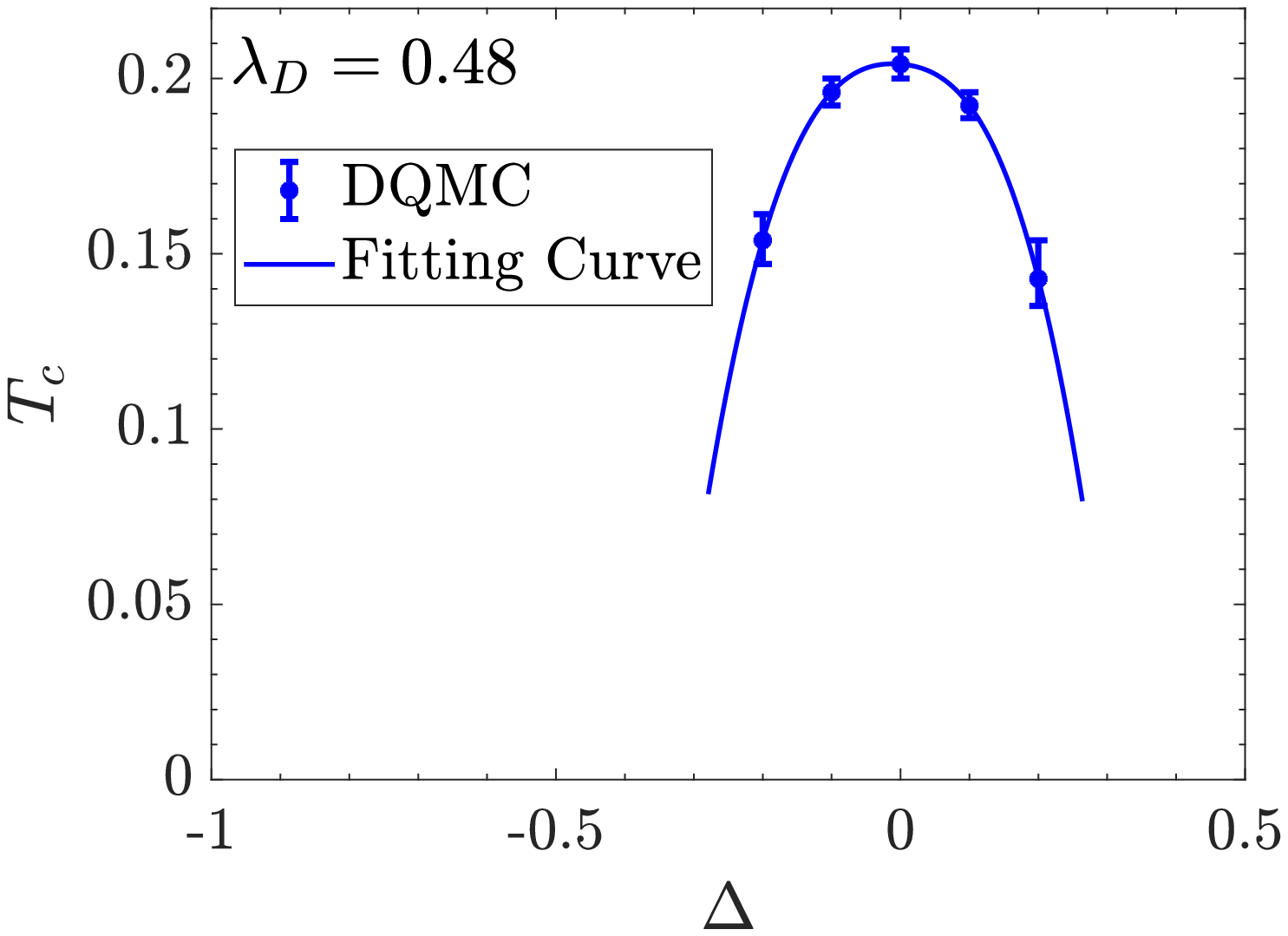}
\includegraphics[height=6.0cm,width=8.0cm]{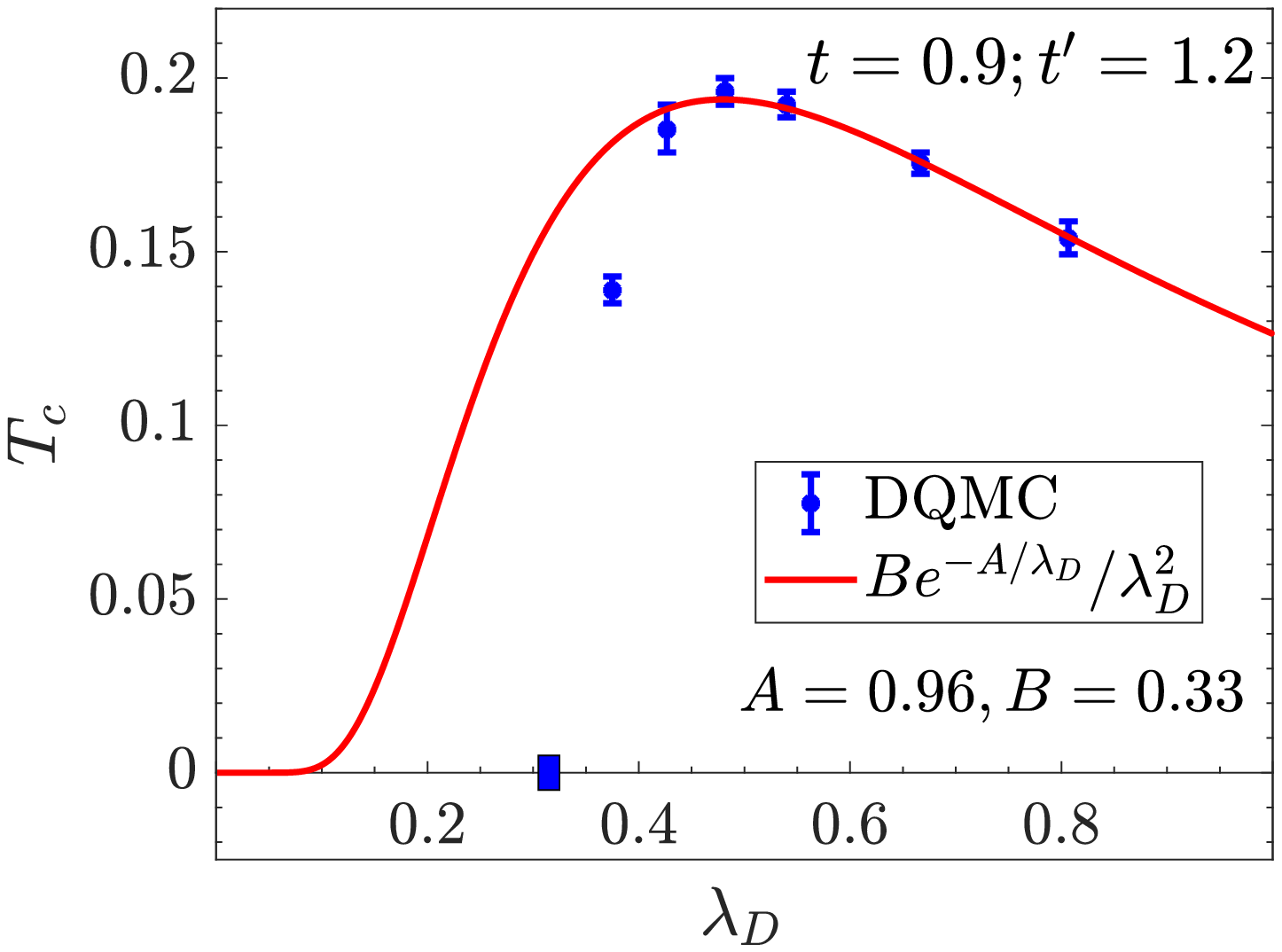}
\caption{
Top: Phase diagram of $T_c$ as a function of $\Delta$ when 
$\lambda_D=0.48$. 
$T_c$ reaches its maximum for isotropic hopping ($\Delta=0$), and 
drops sharply on the dimer side $\Delta<0$ and the hexagon side $(\Delta>0$). 
Bottom: Phase diagram of $T_c$ as a function of $\lambda_D$ 
when $t=0.9,t'=1.2$ with 
$(\lambda_D)_c\sim0.32$.
As $\lambda_D$ grows,   $T_c$ increases first, as the 
electron-phonon coupling induces the CDW phase, but then decreases 
as large values of the electron-phonon coupling cause the polarons
to become increasingly heavy \cite{zhang19,chen19}.
The symbol along the horizontal axis of panel (b) is obtained 
by extrapolating the sharp descent of the DQMC data for $T_c$, combined
with low temperature simulations which show the charge correlations
are short ranged.
}
\label{fig:Tc_Delta_phase_diagram}
\end{figure}

Phase diagrams are obtained for fixed 
$\lambda_D=0.48$, varying $t$ and
$t'$ (top panel) and fixed $t=0.9,t'=1.2$, varying $\lambda_D$ (bottom
panel) in Fig.~\ref{fig:Tc_Delta_phase_diagram}. 
It is numerically challenging to attempt to extract $T_c$ when it
becomes too small.  Nevertheless, we can put reliable upper bounds on 
$T_c$ by measuring $S_{\rm cdw}$ at large $\beta$ and verifying its
value is consistent with only short range charge correlations.
Doing simulations at $\beta$ up to $\beta = 25$ (temperature
$T=0.04$) strongly suggests that,
similar to the
undecorated honeycomb case \cite{zhang19,chen19}, 
there is a nonzero
critical coupling $(\lambda_D)_c\sim0.32$,
for $t=0.9,t'=1.2$,  as
indicated along the $T=0$ axis in the bottom panel,
Fig.~\ref{fig:Tc_Delta_phase_diagram}(bottom). 
Correspondingly, for fixed $\lambda_D=0.48$, large $\beta$ simulations
suggest there are 
critical hopping differences $\Delta_c$, as shown
in the top panel, Fig.~\ref{fig:Tc_Delta_phase_diagram}a. 
The presence of these QCP is further supported by their
appearance in the MFT results in Fig.~\ref{fig:MFT_L4lambda2}.

The fitting curve and estimated QCP in 
Fig.~\ref{fig:Tc_Delta_phase_diagram}(bottom) is based on a physically
motivated form for $T_c$ which combines the weak coupling BCS and a strong
coupling expectation that $T_c$ should decline with large
$\lambda_D$ as the effective electron hopping decreases.  
The curve in 
Fig.~\ref{fig:Tc_Delta_phase_diagram}(top) is based on a simple
cubic spline through the data.  Since this is based on an
{\it ad-hoc} functional form, there is considerable uncertainty in the
positions of the QCPs which are seen in the MFT treatment
(Fig.~\ref{fig:MFT_L4lambda2}).

\begin{figure}[t]
\includegraphics[height=6.0cm,width=8.0cm]{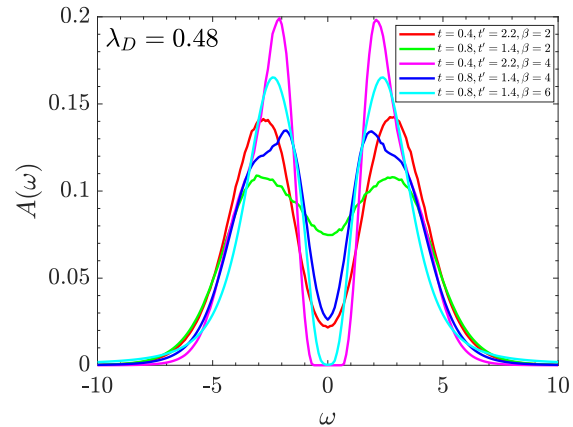}
\caption{
Spectral function for the two cases $\Delta=-0.6$ in the charge liquid
phase and $\Delta=-0.2$ in the CDW phase.  
}
\label{fig:energy_spectralfunction}
\end{figure}

We conclude by examining the single particle spectral function,
$A(\omega)$, which is related to the fermion Greens function
$G(\tau)$ obtained in DQMC {\it via},
\begin{align}
G(\tau) = \int d\omega \frac{e^{-\omega \tau}}{e^{\beta \omega}+1}
\, A(\omega) \,\,.
\label{eq:Aw}
\end{align}
$A(\omega)$ is the many-body analog of the single-particle
density of states, and hence carries information 
concerning the opening of energy gaps in the excitation spectrum.
We invert Eq.~\ref{eq:Aw} via the maximum entropy 
method\cite{gubernatis91}.
Fig.~\ref{fig:energy_spectralfunction} gives
$A(\omega)$ for two values of
hopping difference on opposite sides of the CDW-charge
singlet QCP.  Despite the difference in the nature of the ground state,
$A(\omega)$ vanishes at the Fermi surface $\omega=0$ in both cases, as
$T$ is lowered.  In the case of larger $\Delta$, this
reflects the presence of a charge singlet gap.  In the case of smaller
$\Delta$, this is a CDW gap.  Similar behavior occurs
on the two sides of the antiferromagnetic-spin singlet QCP in the
multiband Hubbard model \cite{euverte13}.

\section{5.  Conclusions}

In this manuscript we have presented Determinant Quantum Monte Carlo
results for the Holstein model with modulated hopping on a `decorated
honeycomb lattice' which consists of a collection of weakly coupled
hexagons, or, in the opposite limit of the relative hybridizations,
weakly coupled dimers.  Our key result was the determination of the
evolution of the charge density wave order as one moves away from
uniform hopping towards either of these extremes.  This work represents
an extension of investigations of the competition between magnetically
ordered and spin liquid phases in decorated Hubbard Hamiltonians, to CDW
to charge singlet transitions in electron-phonon models.  The effect of
$t_{x,y} =(1 \pm \Delta)$ on $S_{\rm cdw}$ has also been recently
studied in the anisotropic square lattice Holstein
Hamiltonian \cite{cohenstead19}.  However, in this case the modulation
endpoints $\Delta=-1,+1$ are decoupled, but still infinite, linear
chains.  In the present work the endpoints $\Delta=-1,+0.5$ result in
small independent clusters.  As a result of infinite clusters still
being present, long range order is somewhat more robust to modulation in
the square lattice case.

The geometry we investigated has been proposed as a possible realization
of a ${\cal Z}_2$ topological state associated with the `artificial
orbitals' of the independent hexagons.  As discussed in \cite{wu15}, it
might be possible to implement this geometry via the placement of a
triangular lattice of CO molecules on a Cu(111) surface.  Our work has
shown that in addition to topological properties, electron-phonon
interactions can show a diverse set of charge ordering behavior on such
lattices.

The strong breaking of the pairing-charge degeneracy distinguishes the
present work from previous magnetic analogs.  Specifically, what we
demonstrate here is that despite the lack of `rotational' symmetry,
local objects which have (imperfect) singlet character nevertheless
still form on the strong bonds, and these ultimately lead to a loss of
CDW order.  This non-trivial result could not be anticipated by magnetic
analogs where rotational symmetry is always exact.
\color{black}
Indeed, we have provided, for the first time to our knowledge, 
a precise quantification of the Holstein to Hubbard mapping in
the anti-adiabatic limit.
The data of Fig.~\ref{fig:dimer} emphasize that
the spin symmetry characterizing
the Hubbard model is violated by more than a factor of five
for the Holstein model at $\omega_0/t=1$, by almost a factor
of two at $\omega_0/t=10$, and even at 
$\omega_0/t=10^2$ a difference of 5 percent remains.

\color{black}

\section{Acknowledgements}

CHF and RTS were supported by Department of Energy grant
DE-SC0014671.
HG was supported by the NSFC grant No.~11774019.


\end{document}